\documentclass[fleqn,twoside,twocolumn,nofootinbib,showkeys]{revtex4} 
\usepackage[sec,nocpr]{ujp} 

\usepackage{graphicx}
\usepackage{dcolumn}
\usepackage{bm}
\usepackage{amsmath}
\usepackage[english]{babel}

\makeatletter\AtBeginDocument{\let\@elt\relax}\makeatother

\begin{document}
\title[Statistical Multifragmentation Model within the Extended Morphological Thermodynamics Approach]
{Statistical Multifragmentation Model within the Extended Morphological Thermodynamics Approach}%

\author{Kucherenko V. S.}
\affiliation{Igor Sikorsky Kyiv Polytechnic Institute, Institute of Physics and Technology}
\address{Peremohy Ave. 37, 03056 Kyiv, Ukraine}
\email{kucherenko.slavaslava@gmail.com}

\author{Bugaev K. A.}
\affiliation{Bogolyubov Institute for Theoretical Physics of the National Academy of Sciences of Ukraine}
\address{Metrolohichna str. 14-b, 03680 Kyiv, Ukraine}
\affiliation{Department of Physics, Taras Shevchenko National University of Kyiv}
\address{Hlushkova Ave. 4, 03127 Kyiv, Ukraine}
\email{Bugaev@th.physik.uni-frankfurt.de}

\author{Sagun V.}
\affiliation{CFisUC, Department of Physics, University of Coimbra}
\address{Rua Larga 3004-516, Coimbra, Portugal}
\email{violetta.sagun@uc.pt}

\author{Ivanytskyi O.}
\affiliation{Institute of Theoretical Physics, University of Wroclaw}
\address{pl. M. Borna 9, 50-204 Wroclaw, Poland}
\email{oleksii.ivanytskyi@uwr.edu.pl}

\udk{№ УДК/UDC} \razd{\secix}

\autorcol{V.S.\hspace*{0.7mm}Kucherenko, K.A.\hspace*{0.7mm}Bugaev, V.\hspace*{0.7mm}Sagun et al.}%

\setcounter{page}{1}%

\begin{abstract}
On the basis of morphological thermodynamics we develop an exactly solvable version of statistical mutifragmentation model for 
the nuclear liquid-gas phase transition. It is shown that the hard-core repulsion between spherical nuclei  generates
only the bulk (volume), surface and curvature parts of the free energy of the nucleus, while the Gaussian curvature one does not appear in the derivation. The phase diagram of nuclear liquid-gas phase transition is studied for  a truncated version of  the  developed model. \\

\end{abstract}

\keywords{morphological thermodynamics, induced surface and surface tensions, equation of state, nuclear liquid-gas phase transition, 
statistical multifragmentation}

\maketitle

\section*{Introduction}
Over the past 20 years the  morphological thermodynamics approach has been extensively developed  in condense matter physics to describe the behavior of dense 3- and 2-dimensional fluids \cite{Kucherenko:Ref1}. It is based on  the Hadwiger theorem \cite{Kucherenko:Ref2,Kucherenko:Ref3}
 and can be formulated as:  
the total  free energy decrease $-\Delta \Omega$ of  a convex rigid body $\textbf{\emph{B}} $ inserted into a fluid 
can be completely described by four thermodynamic characteristics such as the system  pressure  $p$, the mean surface tension coefficient $\Sigma$, the mean  curvature tension coefficient $K$ and the bending rigidity
coefficient $\Psi$, i.e. 
\begin{equation}\label{Eq1}
	-\Delta \Omega = p v_{\emph{B}} +\Sigma s_{\emph{B}} + K c_{\emph{B}} + \Psi x_{\emph{B}}\, ,
\end{equation}
 where the quantities $v_{\emph{B}}, s_{\emph{B}}, c_{\emph{B}}, x_{\emph{B}}$ are, respectively, the volume of 
 $\textbf{\emph{B}}$, its surface, the  mean curvature integrated
over the surface of  rigid body and the  mean Gaussian curvature also integrated over the surface of   $\textbf{\emph{B}}$. The last two quantities are  defined  via the two local principal curvature radii $R_{c1}$ and $R_{c2}$ as
 $c_{\emph B} = \int\limits_{\partial {\cal B}} d^2 r \frac{1}{2} \left[\frac{1}{R_{c1}}+ \frac{1}{R_{c2}}\right] $ and $x_{\emph B} = \int\limits_{\partial {\cal B}} d^2 r \frac{1}{R_{c1} R_{c2}} $ (the Euler characteristic).
Such a treatment is usually justified, if the system exists  relatively away from the critical point and, simultaneously, from wetting and drying transitions \cite{Kucherenko:Ref1,Kucherenko:Ref3}. 

Independently to  the morphological thermodynamics, its analogue in the grand canonical ensemble widely known as the induced surface and curvature tensions (ISCT) equation of state (EoS) was developed recently in  \cite{Kucherenko:Ref4,Kucherenko:Ref5,Kucherenko:Ref6,Kucherenko:Ref7,Kucherenko:Ref8}. This cutting-edge approach was successfully applied in Ref. \cite{Kucherenko:Ref4} to model the properties of  the one and two-component mixtures of classical hard spheres and hard discs. The ISCT  EoS was developed not only for classical particles,  but also  for quantum ones \cite{Kucherenko:Ref5,Kucherenko:Ref6} and for relativistic particles that experience the Lorentz contraction \cite{Kucherenko:Ref8}.
Moreover, very recently the grand canonical ensemble formulation of  morphological thermodynamics was also worked out for very small system volumes of about 100 fm$^3$, which are typical for the nuclear reactions \cite{Kucherenko:Ref7}.

Here we  reformulate  an  exactly solvable version of the  statistical multifragmentation model (SMM) \cite{Kucherenko:Ref9,Kucherenko:Ref10} using the postulates of morphological thermodynamics. 
This is a highly nontrivial extension of the morphological thermodynamics to an infinite number of degrees of freedom which are the nuclear clusters of $k$ nucleons with $k= 1, 2, 3,...$. Besides, this is an important  extension of the usual 
morphological thermodynamics to a new domain, namely to the vicinity of (tri)critical endpoint of nuclear matter.  In addition, here we discuss the possible extension of the suggested model which will improve the description of liquid  phase of nuclear matter at high packing fractions. 

The work is organized as follows. In Sect. 1 we heuristically derive the ISCT EoS for the SMM using the requirements of morphological thermodynamics. In Sect. 2 a truncated version of the ISCT EoS for the SMM is discussed, while our results are summarized in Conclusions.


\section{Derivation of  ISCT EoS for SMM}

%

 We start our discussion  from the  one-component gas of Boltzmann particles  with hard-core repulsion.  The  potential energy $U(r)$ between such  particles depends on the distance $r$ between their centers as
\begin{equation}\label{Eq2}
	U(r) = \left\{ \begin{array}{rcl}
         \infty & \mbox{for}
         & r\leq 2 R \\ 0  & \mbox{for} & r> 2 R \end{array}\right. \,,
\end{equation}
and, hence, the  quantity $R$ is the hard-core radius. Since the typical temperatures of  the nuclear liquid-gas phase transition (PT) 
are below 20 MeV, i.e. much smaller than the nucleon mass,  one can safely use the non-relativistic treatment \cite{Kucherenko:Ref9,Kucherenko:Ref10}. 

In the grand canonical ensemble the  pressure of  Van der Waals (VdW)  EoS of particles with hard-core repulsion can be written as
\begin{eqnarray}\label{Eq3}
	p &=& T \phi(T) \, \exp \left( \frac{\mu-a\, p}{T} \right), \\ 
	\phi(T) &=& g \int \frac{d^3 k}{ (2\pi \hbar)^3}\, \exp \left(-\frac{\textbf{k}^2}{2mT} \right) \, , 
	\label{Eq4n}
\end{eqnarray}
where
$T$ is the system temperature and $\mu$ is its chemical potential.
In  Eq. (\ref{Eq3})  the parameter $a = \int d^3r \left[ 1- \exp\left( -U(r)/T \right) \right] \equiv 4 v \equiv v+  s R \equiv v+ \frac{1}{2} s R +  \frac{1}{2}c R^2$ denotes the second virial coefficient, where  $v=\frac{4}{3}\pi R^3$ denotes the eigenvolume  of particles, while  $s=4 \pi R^2$ and $c= 4 \pi R$ denote, respectively,  their eigensurface and double eigenperimeter. In Eq. (\ref{Eq4n})  $\phi(T)$ is  the thermal density of particles with mass $m$ and the degeneracy factor $g$. 

According to  morphological thermodynamics \cite{Kucherenko:Ref1,Kucherenko:Ref2,Kucherenko:Ref3} the free energy of  rigid particle should be written as $f = v p +s \Sigma + c K$ and, hence, the grand canonical pressure (\ref{Eq3}) should be generalized
as 
\begin{eqnarray}\label{Eq5}
	p = T \phi(T) \, \exp \left( \frac{\mu- (v p +s \Sigma + c K)}{T} \right) \,, 
\end{eqnarray}
but this equation should be supplemented by the equations for the surface tension coefficient $\Sigma$ and the curvature tension one
 $K$ induced by hard-core repulsion. In Ref. \cite{Kucherenko:Ref4} one can find how  the equations for  $\Sigma$ and  $K$ can be derived rigorously. 
 Here we generalize the heuristic derivation of Ref. \cite{Kucherenko:Ref10} by including into our  treatment the  curvature tension
  coefficient $K$.

Since in the SMM the nuclear clusters can consist of any positive number of nucleons, we   consider the  system of $N$-sorts particles of the hard-core radii $R_k$, with $k=1,2,...,N \rightarrow \infty$. 
The virial expansion of the gas pressure up to the second order  can be cast as \cite{Kucherenko:Ref10}
\begin{eqnarray}\label{Eq6n}
p &=& T\sum_{k=1}^{N}\phi_k  e^{\frac{\mu_k}{T}} \left[1- \sum_{n=1}^{N}a_{kn}\phi_n e^{\frac{\mu_n}{T}} \right] ,~ \\
\phi_n(T) &=& g_n\hspace{-1.1mm} \int \hspace{-1.1mm}\frac{d^3 k}{(2\pi \hbar)^3}\hspace{-0.55mm} \exp\left[-\frac{\textbf{k}^2}{2m_nT} \right] = g_n \hspace{-1.1mm}\left[ \frac{m_n T}{2 \pi \hbar^2} \right]^\frac{3}{2}\hspace{-1.1mm},~~~~~
\label{Eq7n}
\end{eqnarray}
where  $\phi_n(T)$ is  the thermal density of particles of the degeneracy $g_n$, mass $m_n=n m_1$ with $m_1=940$ MeV.
The second virial coefficient $a_{kn}$  which has the meaning of excluded volume per particle 
is given by 
\begin{eqnarray}\label{Eq8n}
 a_{kn} &=& \frac{2}{3}\pi (R_k +R_n)^3 =
\nonumber \\
&=& 
\frac{2}{3}\pi(R_k^3+3R_k^2 R_n+ 3R_k R_n^2 +R_n^3) \,. 
\end{eqnarray}

Consider first the low densities at which the expansion  (\ref{Eq6n}) is valid. 
Substituting the second virial coefficients  (\ref{Eq8n}) into Eq.  (\ref{Eq6n}) and regrouping the terms with the same powers of   $R_k$ of a $k$-nucleon fragment, we can write
\begin{eqnarray}\label{Eq9n}
&&p =T\sum_{k=1}^{N}\phi_k  e^{\frac{\mu_k}{T}} \left[1- \frac{4}{3}\pi R_k^3\sum_{n=1}^N\phi_n  e^{\frac{\mu_n}{T}}- \right. \quad \nonumber \\
&&\left. -   \frac{4\pi R_k^2}{2}\sum_{n=1}^N R_n \phi_n  e^{\frac{\mu_n}{T}} -\frac{4\pi R_k}{2} \sum_{n=1}^N R_n^2\phi_n  e^{\frac{\mu_n}{T}} \right] . \quad~
\end{eqnarray}
Apparently, for low densities each  sum  in Eq. (\ref {Eq9n}) can be identified as 
\begin{eqnarray}\label{Eq10n}
\sum_{n=1}^N\phi_n  e^{\frac{\mu_n}{T}} &=& \frac{p}{T}\, , \quad \\
\frac{1}{2}\sum_{n=1}^N R_n \phi_n  e^{\frac{\mu_n}{T}} &=& \frac{\Sigma}{T} \,, \quad 
 \frac{1}{2} \sum_{n=1}^N R_n^2\phi_n  e^{\frac{\mu_n}{T}} = \frac{ K }{T} \,. ~~~\quad
 \label{Eq11n}
\end{eqnarray}
The approximation  (\ref{Eq10n}) is valid for the low densities, since defining the second virial coefficient
with the help of Eqs. (\ref{Eq10n}) and (\ref{Eq11n}),  one
modifies  the third and higher virial coefficients which at these densities can be neglected.  Therefore, substituting Eqs. 
(\ref{Eq10n}) and (\ref{Eq11n}) into the right hand side of Eq. (\ref {Eq9n}), we obtain 
\begin{eqnarray}\label{Eq12n}
&&p =T\sum_{k=1}^{N}\phi_k  e^{\frac{\mu_k}{T}} 
\times \nonumber \\
&& \times  
\left[{\textstyle 1- \frac{4}{3}\pi R_k^3
\frac{p}{T} -   {4\pi R_k^2} \frac{\Sigma}{T} -  {4\pi R_k} \frac{K}{T} }\right]  \simeq \quad  \nonumber \\ 
&& \simeq T\sum_{k=1}^{N}\phi_k  \exp\left[ {\frac{\mu_k - v_k p - s_k\Sigma - c_k K}{T}} \right] \, , ~
\label{Eq13n}
\end{eqnarray}
where in the last step of derivation we used the approximation $1-x \simeq \exp (-x) $ which is valid for low densities. 
In Eq. (\ref{Eq13n}) we introduced the eigenvolume  $v_k = \frac{4}{3}\pi R_k^3$, 
the eigensurface $s_k = 4\pi R_k^2$ and
double eigenperimeter  $c_k = 4\pi R_k$ of the $k$-nucleon cluster with the  hard-core radius $R_k = R_1 k^\frac{1}{3}$ and $R_1 \simeq 0.72$ fm. 

Comparing  Eqs. (\ref {Eq5}) and (\ref {Eq13n}), we conclude that the latter one is a multicomponent 
version of the grand canonical pressure obtained in accord with the morphological thermodynamics. Using the same logic,
we can generalize the  expressions (\ref {Eq11n}) for the coefficients of induced surface tension $\Sigma$ and induced curvature tension $K$ in accordance with the morphological thermodynamics and write
\begin{eqnarray}\label{Eq14n}
\Sigma  &=&  T A\sum_{k=1}^{N} R_k \, \phi_k  \exp\left[ {\frac{\mu_k - v_k p - s_k\Sigma - c_k K}{T}} \right], ~~~\quad  \\
K  &=&  T B \sum_{k=1}^{N} R_k^2 \, \phi_k  \exp\left[ {\frac{\mu_k - v_k p - s_k\Sigma - c_k K}{T}} \right], ~~~
\label{Eq15n}
\end{eqnarray}
where the coefficients $A=0.5$ and $B=0.5$ later on  will be considered as the adjustable parameters. 
In Refs. \cite{Kucherenko:Ref4,Kucherenko:Ref5,Kucherenko:Ref6} it is shown that the system (\ref {Eq13n})-(\ref {Eq15n})
is the VdW EoS  which can be improved at high densities by inserting  the set of parameters $\{ \alpha_k > 1\}$ and $\{\beta_k>1\}$
for each sort of particles into Eqs. for  $\Sigma$ and $K$ as 
\begin{eqnarray}\label{Eq16n}
\frac{\Sigma}{T} \hspace{-0.55mm} &=&  \hspace{-0.55mm}  A\hspace{-1.1mm}\sum_{k=1}^{N} \hspace{-1.1mm} R_k  \phi_k  \exp \hspace{-1.1mm}
\left[ {\frac{\mu_k - v_k p - \alpha_k s_k\Sigma - c_k K}{T}} \right] \hspace{-1.1mm} , \quad  \\
\frac{K}{T} \hspace{-0.55mm}  &=&   \hspace{-0.55mm} B \hspace{-1.1mm}\sum_{k=1}^{N} \hspace{-1.1mm} R_k^2 \phi_k  \exp\hspace{-1.1mm}\left[ {\frac{\mu_k - v_k p - \alpha_k s_k\Sigma - \beta_k c_k K}{T}} \right] \hspace{-1.1mm} , ~~~\quad
\label{Eq17n}
\end{eqnarray}
which can be now extrapolated to any particle number densities.
The principal difference of the obtained system (\ref {Eq13n})-(\ref {Eq15n}) from the ones analyzed previously in 
Refs. \cite{Kucherenko:Ref4,Kucherenko:Ref5,Kucherenko:Ref6} is that the degeneracy factor $g_k$ of $k$-nucleon cluster 
is a statistical partition of the ensemble of clusters with the same mean volume $v_k$, but different shapes  \cite{Kucherenko:Ref10,Kucherenko:Ref11,Kucherenko:Ref12}. For  $k\gg 1$ this internal partition of 
large clusters  of $k$-nucleons can be expressed 
in terms of the mean surface $s_k$  \cite{Kucherenko:Ref10,Kucherenko:Ref11,Kucherenko:Ref12}. Here we
generalize the  internal partition $g_k$ to high pressures  by including the mean double perimeter $c_k$   in accord  with the morphological thermodynamics and write it  as
\begin{eqnarray}\label{Eq18n}
g_{k\gg1} =   \frac{1}{k^{\tau+\frac{3}{2}}} \exp
\left[ {\frac{ v_k p_L -  s_k\sigma_0(T) - c_k K_0 (T)}{T}} \right] \hspace{-1.1mm} , \quad 
\end{eqnarray}
where $p_L$ is the pressure of nuclear liquid (internal pressure of large custers), $\sigma_0(T)$  is the $T$ dependent proper surface tension coefficient and $K_0 (T)$ is
 proper curvature tension coefficient, while $\tau$ is the Fisher exponent  \cite{Kucherenko:Ref11}. Eq. (\ref{Eq18n}) is a generalization of  the expression found in Ref. \cite{Kucherenko:Ref10} which allows us to account for the effects of proper curvature tension.  It is necessary to stress 
 that without  inclusion of the proper curvature tension coefficient in Eq. (\ref{Eq18n}) it is impossible to compensate the induced surface tension one and, hence, in such a model the (tri)critical point does not exist.

Existence of the curvature and even of the Gaussian curvature terms  in the Bethe-Weizs\"acker  formula for the binding energy of
large nuclei at zero temperature  is discussed for about fifty years \cite{Kucherenko:Ref13,Kucherenko:Ref14,Kucherenko:Ref15}, but, so far,    there was no definite conclusion  reached about its existence. In this work we derived the induced curvature tension coefficient similarly to the surface one, but our analysis shows no  room  to introduce the Gaussian curvature terms. Therefore, we  do not consider it for the SMM. 

The $T$-dependence of both the proper and the induced surface tension coefficients of nuclear clusters are discussed for a few decades \cite{Kucherenko:Ref9,Kucherenko:Ref12,Kucherenko:Ref14,Kucherenko:Ref15,Kucherenko:Ref16}. Although there exist several 
parameterizations for $\sigma_0(T)$ and for $\sigma_{tot}=\sigma_0(T)+\Sigma$, we use the linear $T$-dependence of $\sigma_0(T)$, since  it is obtained within an exactly solvable model for the surface partition \cite{Kucherenko:Ref12}. Besides,  a thorough
analysis of experimental data performed in Ref. \cite{Kucherenko:Ref16} shows that there is a wide range of temperatures at which 
the total surface tension coefficient $\sigma_{tot}$ is a linear function of $T$.  Below it is shown that even a truncated version
of the ISCT EoS provides an existence of two regions of such a $T$-dependence of $\sigma_{tot}$.

\section{Truncated ISCT EoS  for low densities} 

The system (\ref {Eq13n})-(\ref {Eq15n})  can be used to describe the properties of nuclear matter at high particle number densities 
and to clarify a principal question what is the value of total surface tension coefficient $\sigma_{tot}$ at supercritical temperatures. In the famous Fisher droplet model \cite{Kucherenko:Ref11} and in the solvable version of SMM \cite{Kucherenko:Ref9}
it is assumed that  $\sigma_{tot} = 0$ for $T \ge T_{cep}$, while in the SMM with compressible nuclear liquid 
it is argued that  $\sigma_{tot} < 0$ for $T>T_{cep}$, while $\sigma_{tot} =0$ at $T=T_{cep}$. Apparently,  this problem can be solved  only experimentally.   
On the other hand, it is expected that the critical point of nuclear matter is located at particle number  densities  about  $\rho_c \simeq \frac{\rho_0}{3}$ \cite{Kucherenko:Ref10,Kucherenko:Ref14}, where $\rho_0 \simeq 0.16$ fm$^{-3}$ is the normal nuclear density. 
Therefore, to study the properties of nuclear matter at the vicinity of critical endpoint  it is sufficient to work out a simpler EoS.

Using the fact that at low particle number densities the contributions of surface and curvature tensions to the second virial coefficient 
are the same, i.e. $\sum_{k=1}^{N}\phi_k  e^{\frac{\mu_k}{T}}R_k^2 \Sigma = \sum_{k=1}^{N}\phi_k  e^{\frac{\mu_k}{T}}R_k K $,  one can account for the curvature tension effects by doubling down the contribution of  $ \Sigma$ on the right hand side of  Eqs. (\ref {Eq13n}) and (\ref {Eq16n})
\cite{Kucherenko:Ref10}. 
Hence, instead of the system (\ref {Eq13n})-(\ref {Eq15n}) one can write
\begin{eqnarray}\label{Eq19n}
&&p = T \sum_{k=1}^N \phi_k \exp \left[\frac{\mu_k - v_k p - s_k \Sigma}{T} \right] ,\\
&&\Sigma = T \sum_{k=1}^N R_k \phi_k \exp \left[\frac{\mu_k - v_k p - \alpha s_k \Sigma}{T} \right] , ~
 \label{Eq20n}
\end{eqnarray}
where the parameter $\alpha =1.5$ is chosen according to Ref. \cite{Kucherenko:Ref10}. 
To parameterize the degeneracy of large clusters according to Eq. (\ref{Eq18n}) and to account for the fact that nucleons have no 
proper surface tension, we assume the thermal density of $k$-nucleon fragments to be as
\begin{eqnarray}\label{Eq21n}
 \phi_1 &=& z_1\left[\frac{mT}{2\pi \hbar^2}\right]^{\frac{3}{2}} \exp\left[-\frac{\sigma_0 (T)}{T}\right]  \\
 \phi_{k \geq 2} &=&  \frac{1}{k^\tau} \left [\frac{m T}{2 \pi \hbar^2 } \right]^{\frac{3}{2}} 
  \exp\left[{\textstyle \frac{(k p_L V_1 - \mu_k) - \sigma_0 (T) k^\frac{2}{3} } {T}}  \right],~~~~~
\label{Eq22n}
\end{eqnarray}
 where the chemical potential of a k-nucleon fragment is $\mu_k = k \mu$ \cite{Kucherenko:Ref9,Kucherenko:Ref10}, and value of  $\tau =1.9$  provides the existence of 1-st order PT for $T< T_{cep}$ \cite{Kucherenko:Ref10}.
 In Eq.  (\ref{Eq21n}) $z_1=4$ is the degeneracy factor of nucleons. For  $k\ge 1$ one finds $v_k= V_1 k= \frac{4}{3} \pi R_k^3$ with  $V_1 = \frac{1}{\rho_0}$.  For $\phi_{k\geq 2}$ the binding energy of nucleons is included  into the pressure of liquid phase
\begin{equation}\label{Eq23n}
p_L= \frac{\mu+W_F(T)+ W_0+ a_{\nu} \left[ \mu+W_0\right]^{\nu}}{V_1} , 
\end{equation}
${\rm with}~ \nu  = 2,3,4.$
The binding energy per nucleon  is  given by $W(T) = W_0 +W_F(T)$, where $W_0=16$ MeV  is the bulk binding energy per nucleon at $T=0$ and $W_F(T) = \frac{T^2}{\varepsilon_0}$ (with $\varepsilon_0 = 16$ MeV)  accounts for the Fermi motion of nucleons inside a  nucleus at $T> 0$ \cite{Kucherenko:Ref9}.
The constant $a_{\nu}$ in Eq.  (\ref{Eq23n}) should  be found by requiring  that at $T=0$ and normal nuclear density $\rho_L = \frac{\partial p_L}{\partial \mu} = \rho_0$ the liquid pressure is zero \cite{Kucherenko:Ref10}. 
For definiteness, in this work we assume that $a_{2} = 1.261\cdot 10^{-2} \: {\text{MeV}}^{-1}$ and $\nu = 2$.
Hence now the system (\ref {Eq19n}), (\ref {Eq20n}) becomes
\begin{eqnarray}\label{Eq24n}
\hspace*{-0.2cm} \frac{p}{T} \hspace*{-0.2cm} &=& \hspace*{-0.2cm} \left[  \frac{m T}{2\pi \hbar^2}\right]^{\frac{3}{2}} \hspace*{-0.2cm} \sum_{k=1}^N \frac{b_k}{k^\tau} \exp\hspace{-1.1mm}\left[ {\textstyle \frac{(p_L - p) V_1 k-(\Sigma+\sigma_0)k^\frac{2}{3}  }{T} } \right], \\
\hspace*{-0.2cm} \frac{\Sigma}{3 V_1 T}\hspace*{-0.2cm} &=&  \hspace*{-0.2cm} \left[  \frac{m T}{2 \pi \hbar^2}\right]^{\frac{3}{2}} \hspace*{-0.2cm} \sum_{k=1}^N \frac{b_k}{k^{\tau-\frac{1}{3}}} \exp \hspace{-1.1mm}\left[{\textstyle  \frac{(p_L - p) V_1k -(\alpha \Sigma+\sigma_0)k^\frac{2}{3}}{T}} \right],\ ~~ \nonumber
 \\~~~~
\label{Eq25n}
\end{eqnarray}
where the degeneracies $b_k $  are defined as  $b_1  = 4 \exp\left[ \frac{-W(T)}{T}\right]$ and $b_{k>1}  = 1$.  The proper surface tension coefficient $\sigma_0 = \sigma_{01} - \sigma_{02} \frac{T}{T_{cep}}$ is chosen 
according to the exact result  found for the surface partition \cite{Kucherenko:Ref12}, where 
$\sigma_{01} = 18 \: \text{MeV}$,  $\sigma_{02} = 24.76$ MeV, $T_{cep}=18$ MeV. 

\begin{figure}[t]
\includegraphics[width=1\columnwidth]{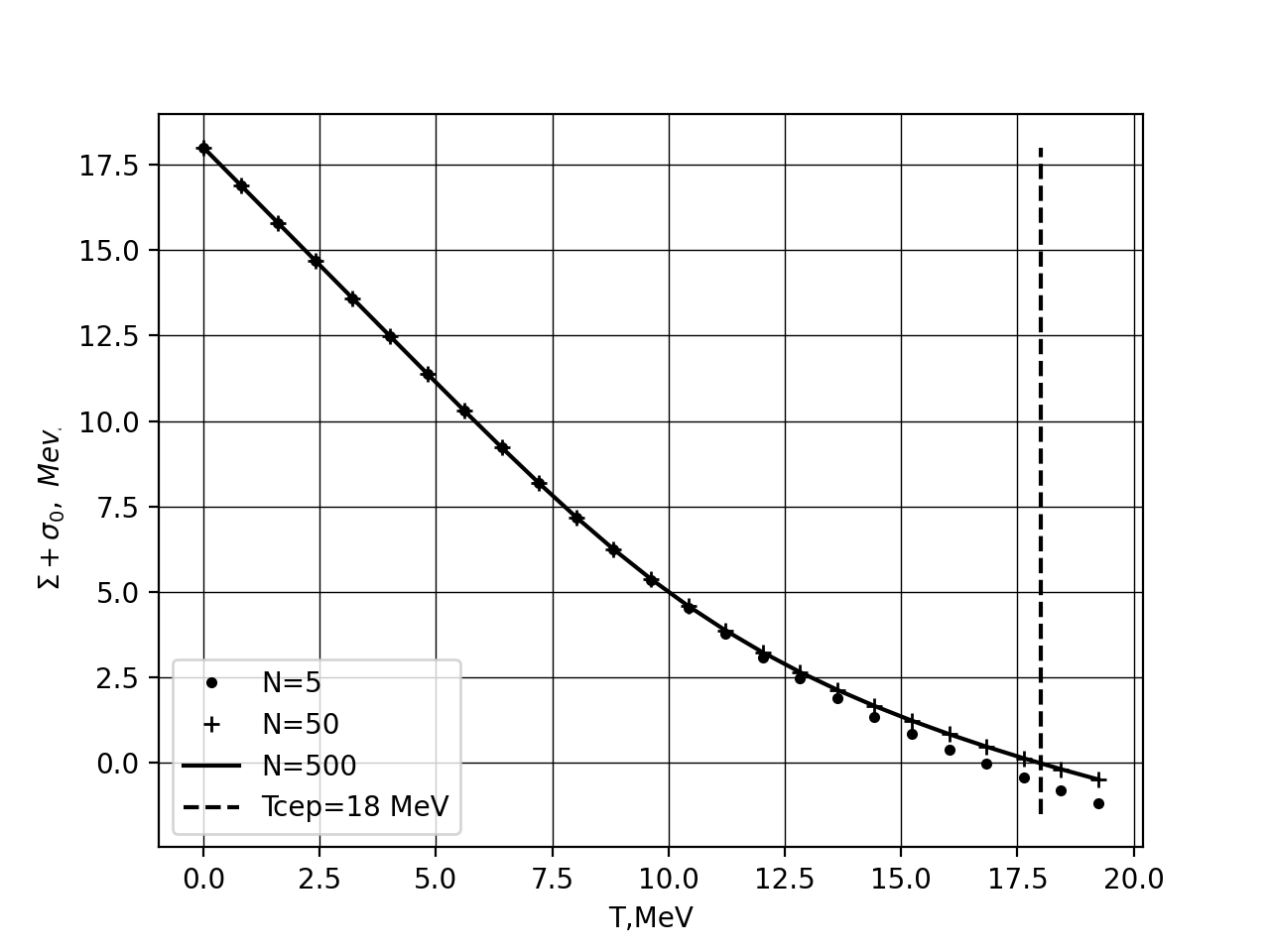}
\caption{Total surface tension coefficient $\sigma_{tot}(\mu_c(T),T)$ at the PT curve $\mu_c(T)$ as a function of $T$ is shown for  several  sizes of largest  nucleus $N$. A vertical line defines the critical temperature $T_{cep}=18$ MeV.}
\label{Kucherenko_fig1}
\end{figure}

Assuming  in Eqs.  (\ref{Eq24n}, \ref{Eq25n}) that $p = p_L$, one obtains  the equations for the PT curve
\begin{eqnarray}\label{Eq26n}
\frac{p_L}{T} \hspace*{-0.2cm} &=& \hspace*{-0.2cm} \left[  \frac{m T}{2\pi \hbar^2}\right]^{\frac{3}{2}} \sum_{k=1}^N \frac{b_k}{k^\tau} \exp\hspace{-1.1mm}\left[ - \frac{(\Sigma+\sigma_0)}{T} k^\frac{2}{3} \right], ~ \\
\label{Eq27n}
\frac{\Sigma}{3 V_1 T} \hspace*{-0.2cm} &=& \hspace*{-0.2cm} \left[  \frac{m T}{2 \pi \hbar^2}\right]^{\frac{3}{2}} \sum_{k=1}^N \frac{b_k}{k^{\tau-\frac{1}{3}}} \exp\hspace{-1.1mm}\left[- \frac{(\alpha \Sigma+\sigma_0)}{T} k^\frac{2}{3} \right].~
\end{eqnarray}

\begin{figure}[t]
\includegraphics[width=1.0\columnwidth]{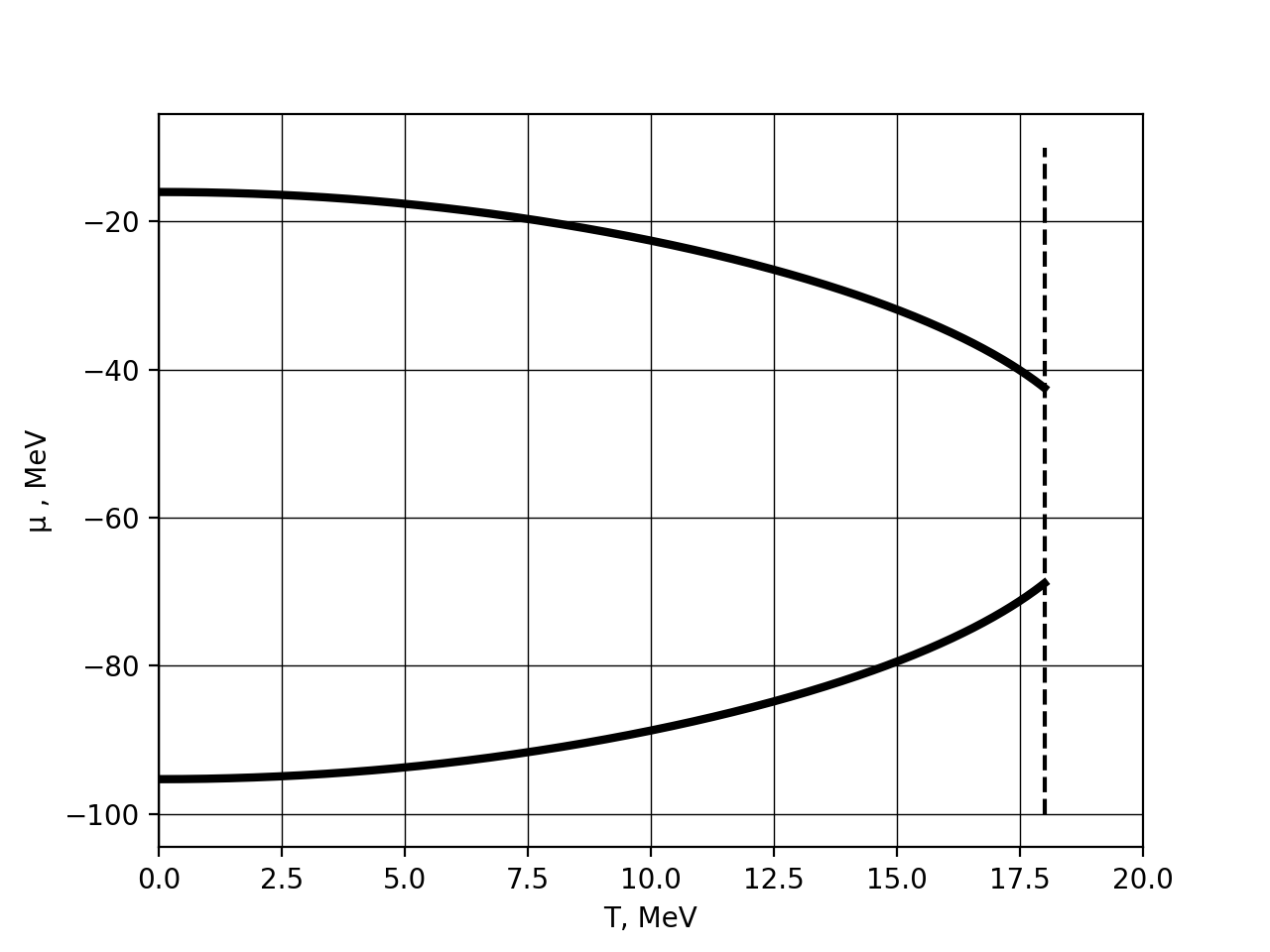}
\caption{The PT curves $\mu_c(T)$ in the plane of temperature $T$ and nuclear chemical potential $\mu$.}
\label{Kucherenko_fig2}
\end{figure}

Solving Eq. (\ref{Eq27n}) for $\Sigma$  and  substituting it  into  Eq. (\ref{Eq26n}),  one gets the  pressure  $p_c = p_L$ at the PT curve.
Then Eq.  (\ref{Eq23n}) for $\mu$  with $\nu=2$ can be written as
\begin{equation}\label{Eq28n}
\mu^2+\mu(1+2 a_2 W_0)+ W(T)+ a_2W_0^2 - p_c V_1 = 0.
\end{equation}
Solving Eq. (\ref{Eq28n}) for  $\mu$, one obtains $\mu_c(T)$ at the PT curve. 
The $T$-dependence of total surface tension coefficient $\sigma_{tot}(\mu_c(T),T)=\Sigma+\sigma_0$ is shown in Fig. \ref{Kucherenko_fig1}, while
the function $\mu_c(T)$  is presented in Fig. \ref{Kucherenko_fig2}. 
From Fig. \ref{Kucherenko_fig1} one can see that $\sigma_{tot}(\mu,T) $ vanishes at $T= T_{cep}=18$ MeV
even for a small size of largest nucleus $N=50$, although the true PT exists only for $N\rightarrow \infty$.

In  Fig. \ref{Kucherenko_fig2} the solutions  $\mu_c(T)$ of Eq.  (\ref{Eq28n}) are shown. 
It is evident that the upper  curve describes the gas-liquid  PT curve
for nuclear matter, while  the lower one corresponds to anti-matter.\\

\bigskip

\section*{Conclusions}
In this work we develop an exactly solvable version of statistical mutifragmentation model for 
nuclear liquid-gas PT  using  the requirements of morphological thermodynamics. 
By evaluating  the second virial coefficients we explicitly demonstrate that the hard-core repulsion between spherical nuclei 
generates only the bulk (volume), surface and curvature parts of the  free energy of nuclei and does not produce  the Gaussian curvature one.
For a   truncated version of  the  developed ISCT EoS we studied the $T-\mu$  phase diagram of nuclear liquid-gas PT for  several sizes of largest  nucleus.

\noindent 
{\bf Acknowledgements.} The present work was partially supported by the National Academy of
Sciences of Ukraine (project No. 0118U003197). V.S. acknowledges the support by national funds from FCT \-- Funda{\c c}{\~a}o para a Ci{\^e}ncia e a Tecnologia, I.P., within the Projects No.  UIDB/04564/2020 and UIDP/04564/2020. The work of O.I. was supported by the Polish National Science Center under the grant No. 2019/33/BST/03059.

\end{document}